\documentclass[12pt]{article}

\newenvironment{pf}{\noindent\textbf{Proof.}\quad}{\hfill{$\Box$}}
\usepackage[colorlinks=false]{hyperref}
\usepackage[dvips]{graphicx}
\usepackage[usenames,dvipsnames]{color}
\usepackage{epsfig}
\usepackage{rotating}
\usepackage{amssymb}
\usepackage{amsmath,amsfonts}
\usepackage[latin1]{inputenc}
\usepackage{verbatim}
\usepackage[tableposition=top,font=small,labelfont=bf,format=hang]{caption}
\usepackage{booktabs}
\newtheorem{lem}{Lemma}
\newtheorem{thm}{Theorem}
\newtheorem{prop}{Proposition}

\newcommand{\qed}{\hfill $\Box$ \\}

\begin{document}

\title{Construction of MDS self-dual codes\\from\\ orthogonal matrices}
\author{
Minjia Shi\thanks{ Minjia Shi, Key Laboratory of Intelligent Computing Signal Processing, Ministry of Education, Anhui University, No.3 Feixi Road, Hefei, Anhui, 230039, China, School of Mathematical Sciences, Anhui University, Hefei, Anhui, 230601,
China and National Mobile Communications Research Laboratory, Southeast University, China, {\tt smjwcl.good@163.com}}, 
Lin Sok\thanks{Lin Sok, School of Mathematical Sciences, Anhui University, Hefei, Anhui, 230601 and Department of Mathematics, Royal University of Phnom Penh, Cambodia, { \tt sok.lin@rupp.edu.kh}}, and
Patrick Sol\'e\thanks{CNRS/LAGA, University of Paris 8, 93 526 Saint-Denis, France , {\tt sole@enst.fr}}
}
\date{}
\maketitle
\begin{abstract}
In this paper\footnote{This work was partly presented at the 3rd Sino-Korea International Conference on Coding Theory and Related Topics, August 12-16, 2016, Beijing, China.}, we give algorithms and methods of construction of self-dual codes over finite fields using orthogonal matrices. Randomization in the orthogonal group, and code
 extension are the main tools. Some optimal, almost MDS, and MDS self-dual codes over both small and large prime fields are constructed.

\end{abstract}

\section{Introduction}

Self-dual codes are one of the most interesting classes of linear codes. They have close connections with group theory, lattice theory, design theory, and modular forms. It is well known that self-dual codes are asymptotically good~\cite{MacSloTho}. Being optimal codes, MDS self-dual codes have been of much interest from many researchers. Optimal self-dual codes over small finite fields were constructed in \cite{GabOtm}. Betsumiya et al. \cite{BetGeoMas} constructed some MDS and almost MDS self-dual codes of length up to $20$ over prime fields ${\mathbb F}_q, 11\le q \le 29$. Georgiou et al. \cite {GeoKou} gave constructions over some larger prime fields up to length $14$. Grassl et al. \cite{GraGul} proved existence of MDS codes of all lengths over ${\mathbb F}_{2^m}$  and of all highest length over finite fields of odd characteristics. Kim et al. \cite{KimLee} studied MDS self-dual codes over Galois rings. Guenda \cite{Gue} constructed MDS Euclidean and Hermitian self-dual codes over larger fields. Jin et al. \cite{JinXin} proved existence of MDS self-codes ${\mathbb F}_q$ in odd characteristic for $q\equiv 1 \pmod 4$ and for $q$ being a square of prime with restricted lengths.

Recently in the classification of extremal binary self-dual codes of length $38$ \cite{AGKSS}, all nonequivalent extremal codes of shorter lengths were reconstructed from the so-called orthogonal matrices and from the method \cite{AG}.

In this paper we generalize the constructions of \cite{AGKSS} from the binary field to arbitrary finite fields. Two main type of constructions are given:
 random sampling from the orthogonal group and code extension by two or four symbols. We focus on large fields and MDS, or near MDS codes. Specifically,
we consider fields of size up to $109,$ and lengths up to $22.$ Over fifty MDS codes with new parameters are constructed. Generator matrices and weight enumerators are archived in  {\tt http://math.ahu.edu.cn/web/user.asp?id=25}
The paper is organized as follows: Section II gives preliminaries and background for self-dual codes. Section III gives method to construct and to extend  a self-dual code. In Section IV we present numerical results of some optimal codes, almost MDS and MDS self-dual codes over different large fields.

\section{Preliminaries}
\label{Sec-Prelim}

We refer to~\cite{HufPle} for basic definitions and results related
to self-dual codes. A {\em
linear $[n,k]_{{\mathbb F}_q}$ code $C$ of length $n$} over ${{\mathbb F}_q}$ is a $k$-dimensional subspace
of  $ {\mathbb F}_q^n$. An element in $C$ is called a {\em codeword}. The
(Hamming) weight wt$({\bf{x}})$ of a vector ${\bf{x}}=(x_1, \dots,
x_n)$ is the number of non-zero coordinates in it. The {\em minimum
distance ({\rm{or}} minimum weight) $d(C)$} of $C$ is
$d(C):=\min\{{\mbox{wt}}({\bf{x}})~|~ {\bf{x}} \in C, {\bf{x}} \ne
{\bf{0}} \}$. The {\em Euclidean inner product} of ${\bf{x}}=(x_1,
\dots, x_n)$ and ${\bf{y}}=(y_1, \dots, y_n)$ in ${\mathbb F}_q^n$ is
${\bf{x}}\cdot{\bf{y}}=\sum_{i=1}^n x_i y_i$. The {\em dual} of $C$,
denoted by $C^{\perp}$ is the set of vectors orthogonal to every
codeword of $C$ under the Euclidean inner product. A linear code
$C$ is called {\em self-orthogonal} if $C\subset C^{\perp}$ and {\em self-dual} if $C=C^{\perp}$. It is well known that a self-dual code can only exist for even length. If $C$ is a self-dual $[2n,n]_{{\mathbb F}_q}$code, then from the Singleton bound, its minimum distance is bounded by
$$d(C)\le n+1.$$
A self-dual code meeting the above bound is called {\em Maximum Distance Separable} ({ MDS}) self-dual code.  A self-dual $[2n,n]$code $C$ is called {\em almost} MDS or {\em near} MDS if $d(C)=n$. A code is called {\it optimal} if it has the highest possible minimum distance for its length and dimension and thus an MDS self-dual code is optimal.

\medskip
In the following, we present some elements used to generate an orthogonal group. In the sequel $\mathbb{F}_q$ denotes a prime field.

The {\em orthogonal group} of index $n$ over a finite field with $q$ elements is defined by
$${\cal O}_n(q):=\{A\in GL(n,q)| AA^T=I_n\}.$$
Let $\theta=\frac{q-1}{2}\in {\mathbb F}_q.$ For a $\{0,1\}-$coordinate vector ${\bf u} \in \mathbb{F}_q^n$ with $wt({\bf u})=4$, define
the $transvection$ determined by ${\bf u}$ as
$$\begin{array}{ll}
t_{{\bf u},\alpha}: &\mathbb{F}_q^n \longrightarrow \mathbb{F}_q^n\\
 &{\bf x} \mapsto
\begin{cases}
 {\bf x}+({\bf x.u}){\bf u},\text{ if } q=2\\
{\bf x}+\theta({\bf x.u}){\bf u},\text{ otherwise.}
\end{cases}
\end{array}
$$
Denote by
$${\cal K}_{\bf u}:=\langle {\cal P}_n ,T_{{\bf u},\theta}\rangle,$$
where ${\cal P}_n$ is the set of $n\times n$ permutation matrices, a subgroup of ${\cal O}_n(q).$ Then the matrix ${T_{{\bf u},\theta}}$ of $t_{{\bf u},\theta}$ in the canonical basis $\{ {\bf e}_1,\hdots,{\bf e}_n \}$ is the symmetric matrix determined by ${T_{{\bf u},\theta}}=I_n+{\bf u}^T{\bf u}$ if $q=2$ and ${T_{{\bf u},\theta}}=I_n+\theta ({\bf u}^T{\bf u})$ otherwise. Moreover ${T_{{\bf u},\theta}^2}=I_n$ and thus ${\cal K}_{\bf u}$ is a subgroup of ${\cal O}_n(q).$

For $q=2$, with the convention ${\cal O}_n:={\cal O}_n(2),$ we have the following theorem due to Janusz \cite{J}.
\begin{thm}[{\cite{J}\label{jan.2}}, Theorem 19]
The orthogonal groups ${\cal O}_n$  are generated as follows
\begin{enumerate}
\item for $1\leq n\leq 3$, ${\cal O}_n={\cal P}_n$,
\item for $n \geq 4$, ${\cal O}_n={\cal K}_{\bf u}$.
\end{enumerate}
\end{thm}
%

For $q=3$, we have the following theorem.
\begin{thm} ${\cal O}_n(3)={\cal K}_{\bf u}$ for $ n\ge 6$.
\end{thm}
To prove the theorem we claim that the orbit ${\bf e}_1{\cal K}_u$ of ${\bf e}_1$ contains all the elements
$\bf w$ such that $||{\bf w}||={\bf w}.{\bf w}^T=1+3s$ for some non-negative integer $s$.\\

\noindent
{\bf Proof of the claim.} We prove by induction on $s$. Note that if $T_{{\bf u},\theta}\in {\cal K}_{\bf u}$, then for any vector ${\bf v}$ obtained from ${\bf u}$ by coordinate permutation, $T_{{\bf v},\theta} \in {\cal K}_{\bf u}$.
Assume that ${\bf u}={\bf e}_1+{\bf e}_2+{\bf e}_3+{\bf e}_4.$
Clearly ${\bf e}_1\in {\bf e}_1{\cal K}_u$ with $||{\bf e}_1||=1$. Assume that the orbit contain all elements $\bf w$ satisfying $||{\bf w}||=1+3s$ for some non-negative integer $s.$ Then
${\bf v}=2{\bf e}_4+{\bf e}_5+\cdots+{\bf e}_{3s+4}\in {\bf e}_1{\cal K}_u$. Now ${\bf x}=2{\bf u}+{\bf v}={\bf v}T_{{\bf u},\theta}\in {\bf e}_1{\cal K}_{\bf u}$ with ${\bf x}.{\bf x}^T=3(s+1)+1$.\qed

\noindent
{ \bf Proof of the theorem.} We prove by induction on $n.$ For $n=6$, $|{\cal K}_{\bf u}|=|{\cal O}_6(3)|= 26127360$ and thus ${\cal K}_{\bf u}={\cal O}_6(3)$. Assume that it is true for $(n-1)\ge 6.$ Let $A\in {\cal O}_n(3).$ Then ${\bf e}_1A\in {\bf e}_1{\cal K}_{\bf u}$ since $||{\bf e}_1A||\equiv 1\pmod 3$  and so ${\bf e}_1AB^{-1}={\bf e}_1$ for some $B\in {\cal K}_{\bf u}.$ Since $AB^{-1}\in {\cal O}_n(3)$, there exists $A_0\in {\cal O}_{n-1}(3)$  such that
$$
AB^{-1}=\left(
\begin{array}{cc}
1&0\\
0&A_0
\end{array}
\right).
$$
Now by induction hypothesis, there exists a vector ${\bf u}_0$ of weight $4$ such that ${\cal O}_{n-1}(3)=\langle {\cal P}_{n-1}, T_{{\bf u}_0,\theta}\rangle.$ It implies that $diag(1,A_0)\in \langle {\cal P}_{n}, T_{{\bf u}_0,\theta}\rangle$. So $AB^{-1}\in {\cal K}_{\bf u}$ and $A\in {\cal K}_{\bf u}.$\qed
\begin{table}\label{Table:1}\caption{Order of ${\cal K}_u$ for $n=5$ }
$$
\begin{array}{|c|c|c|}
\hline
q&|{\cal K}_u|&|{\cal O}_5(q)|\text{\cite{Mac}}\\
\hline
5&18720000&18720000\\
7&276595200&553190400\\
11&51442617600&51442617600\\
13&274075925760&274075925760\\
17&2008994088960&4017988177920\\
19&12228071558400&12228071558400\\
23&41348052472320&82696104944640\\
\hline
\end{array}
$$
\end{table}\\
{\bf Conjecture:} From our computational results, for $q\ge 5$ the group ${\cal K}_{\bf u}$ is either the orthogonal group ${\cal O}_n(q)$ or a subgroup of ${\cal O}_n(q)$ of index $2$ as shown in Table 1.
\section{Construction methods}
In this section, we introduce two constructions of self-dual codes, construction by orthogonal matrices and recursive construction.
\subsection{First construction}
\begin{lem}\label{lem:1} Let $C$ be a linear code of length $2n$ over ${\mathbb F}_q$ with its generator matrix written in the systematic form
$$G_n=\left(
\begin{array}{c|c}
I_n&A\\
\end{array}
\right),$$
where $I_n$ is the identity matrix and $A$ is a matrix of index $n.$
Then $C$ is self-dual if and only if $AA^T=-I_n.$
\end{lem}
{\pf} Since $C$ is self-dual, writing down the parity check matrix $H$ and the generator matrix $G$ in the systematic form and using the fact that $GH^T=0$, we get the result as claimed.\qed

We are now interested in the set $\{A\in GL(n,q): AA^T=-I_n\}$ and for $q=2$, it is exactly the orthogonal group ${\cal O}_n$. With the generation theorem, using the computer software Magma \cite{mag}, we can easily randomly generate binary self-dual codes in large dimension especially the so-called extremal codes.

Note that for $q\neq 2$, $\{A\in GL(n,q): AA^T=-I_n\}$ is not a group any more but a one sided coset of ${\cal O}_n(q).$ If $A$ satisfies $AA^T=-I_n,$ and $B$ is an orthogonal matrix then
$C=BA$ satisfies $CC^T=-I_n.$

 From elementary number theory \cite{IR}, we know that the equation $\alpha^2\equiv -1 \pmod q$ has solutions for $q\equiv 1 \pmod 4$
  (Fermat's two squares theorem), and also $\alpha^2+\beta^2\equiv -1 \pmod q$ has solutions for any prime $q$ (Legendre's three squares theorem). Thus self-dual codes over ${\mathbb F}_q$ can now be constructed as follows.

For lengths that are even and not multiples of $4$, we have the following proposition.
\begin{prop} Let $q\equiv 1 \pmod 4$. Fix $\alpha \in {\mathbb F}_q$ such that $\alpha^2\equiv -1 \pmod q$. Then the matrix $G_n$ of the following form:
\begin{equation}\label{ortho1}
G_n=\left(
\begin{array}{c|c}
I_n & \alpha L\\
\end{array}
\right),
\end{equation}
where $ L\in {\cal O}_n(q)$, generates a self-dual $[2n,n]$ code.
\end{prop}
{\pf The result follows from Lemma \ref{lem:1}.\qed
}

For lengths that are multiples of $4$, we have the following propositions.
\begin{prop}  Fix  $\alpha,\beta \in {\mathbb F}_q$ such that $\alpha^2+\beta^2\equiv -1 \pmod q$ and
$D_0=\left(
\begin{array}{cc}
 \alpha&\beta\\
- \beta&\alpha
\end{array}
 \right)$.
Then the matrix $G_n$ of the following form:
\begin{equation}\label{ortho2}
G_n=\left(
\begin{array}{c|c}
I_{2n} & D_n L\\
\end{array}
\right),
\end{equation}
where $ L\in {\cal O}_{2n}(q), D_n=diag(D_0,\hdots,D_0)$, generates a self-dual $[4n,2n]$ code.
\end{prop}
{\pf The result follows from Lemma \ref{lem:1}.\qed
}
Note that the construction (\ref{ortho2}) is applicable not only for $q\equiv 3 \pmod 4$ but also for $q\equiv 1 \pmod 4$.
\begin{prop} Fix  $\alpha,\beta \in {\mathbb F}_q$ such that $\alpha^2+\beta^2\equiv -1 \pmod q$.
Then the matrix $G_n$ of the following form:
\begin{equation}\label{ortho3}
G_n=\left(
\begin{array}{c|cc}
& \alpha L&\beta L\\
I_{2n}&&\\
& -\beta L^T&\alpha L^T\\
\end{array}
\right),
\end{equation}
where $ L\in {\cal O}_{n}(q),$ generates a self-dual $[4n,2n]$ code.
\end{prop}
{\pf The result follows from Lemma \ref{lem:1}.\qed
}

\begin{prop} Let $C$ be a self-dual $[2n,n]$ code with its generator matrix
$G_n=\left(
\begin{array}{c|c}
I_n&A\\
\end{array}
\right)$. Let $L_1,\hdots, L_r \in {\cal O}_n(q)$. Then the matrix
\begin{equation}\label{eq:conf-diff}
G'_n=\left(
\begin{array}{c|c}
I_n&A L_1 \hdots L_r\\
\end{array}
\right)
\end{equation}
  generates a self-dual $[2n,n]$ code.
\end{prop}
{\pf The result follows from Lemma \ref{lem:1}.\qed
}
{\bf Remark}
In the above construction (\ref{eq:conf-diff}), for $n$ large, in practice $L_1,\hdots,L_r$ are randomly sampled from ${\cal O}_n(q)$. In \cite {GabOtm} Gaborit et al. applied confusion-diffusion rules to the generator matrices of self-dual codes to construct optimal codes. To reduce the number of operations in matrix multiplications, we only apply these rules to the orthogonal-like matrices of the above constructions.

\subsection{Second construction}
Note that if $C_n$ is a linear $[2n,n,d]$ code then $C_n$ can be decomposed as a direct sum
$C_n=D \oplus E$, where $D$ (resp. $E$) is a subcode of $C_n$ of minimum weight $d$ (resp. $e>d$). Moreover the generator matrix $G_n$ of $C_n$ can be written as:
\begin{equation}\label{eq:1}
G_n=\left(
\begin{array}{c}
G_d\\
G_e\\
\end{array}
\right).
\end{equation}
This decomposition allows us to efficiently construct self-dual $[2n+2,n+1,\ge d]$ codes from a self-dual $[2n,n,d]$ code when the dimension of the subcode $D$ is large. For example for $q\equiv 1 \pmod 4,$ a self-orthogonal code of length $2n+2$ can be obained, by extending two coordinates, from a self-dual code of length $2n$ with its generator matrix $G_n$ of the above form (\ref{eq:1}) as follows.
\begin{equation}
\left(\begin{array}{ccccc}
& & &0&0\\
& G_d&&\vdots&\vdots\\
& & &0&0\\
\hline
& & &\lambda_1a&\lambda_1\\
& & &\lambda_2(-1)&\lambda_2a\\
& G_e&&\vdots&\vdots\\
&  & &\lambda_{2i-1}a&\lambda_{2i-1}\\
&  & &\lambda_{2i}(-1)&\lambda_{2i}a\\
& &&\vdots&\vdots\\
\end{array}\right),
\end{equation}
where $a^2\equiv -1 \pmod q$ and $\lambda_1,\lambda_2, \hdots \in {\mathbb F}_q.$

The following propositions construct self-orthogonal $[2n+2,n]$ codes, (resp. $[2n+4,n]$ codes) from a self-dual $[2n,n]$ code, which can later be completed by a direct sum with a one-dimensional, (resp. a two-dimensional subcode) to produce self-dual $[2n+2,n+1]$, (resp. $[2n+4,n+2]$) codes.
\begin{prop}  Let $q\equiv 1 \pmod 4$.  Let $C_n$ be a self-dual code $[2n,n,d]$ over ${\mathbb F}_q$ with its generator matrix $G_n$. Fix $a\in {\mathbb F}_q$ such that $a^2\equiv -1 \pmod q$. Then for any $\lambda_1,\hdots, \lambda_n \in {\mathbb F}_q$, an extended code ${\bar{C}_n}$ of $C_n$ with the following generator matrix $G_{\bar{C}_n}$ is a self-orthogonal $[2n+2,n,\ge d]$ code:
\begin{equation} \label{eq:recursive1}
G_{\bar{C}_n}=\left(\begin{array}{ccccc}
& & &\lambda_1a&\lambda_1\\
& & &\lambda_2(-1)&\lambda_2a\\
& G_n&&\vdots&\vdots\\
&  & &\lambda_{2i-1}a&\lambda_{2i-1}\\
&  & &\lambda_{2i}(-1)&\lambda_{2i}a\\
& &&\vdots&\vdots\\
\end{array}\right).
\end{equation}
\end{prop}
{\pf Since $C_n$ is self-dual,  with the assumption $a^2\equiv -1 \pmod q$, each row of $G_{\bar{C}_n}$, which is an extended row of $G_n$, is orthogonal to itself and to the other rows and thus the result follows.
\qed
\begin{prop}  Let $C_n$ be a self-dual code $[2n,n,d]$ over ${\mathbb F}_q$ with its generator matrix $G_n$. Fix $a,b, c, d \in {\mathbb F}_q$ such that $a^2+b^2+c^2+d^2\equiv 0 \pmod q$.
Let $\bar C_n$ be a code obtained from $C_n$ by extending four coordinates. Let ${\bf x} \in {{\bar C}_{n}}^\perp\slash C_{n} $. Then for any $\lambda_1,\hdots, \lambda_n \in {\mathbb F}_q$, a code ${{C}'_{n}}$ with the following generator matrix $G_{{C}'_{n}}$ is a self-orthogonal $[2n+4,n+1]$ code:
\begin{equation} \label{eq:recursive2}
G_{{C}'_{n}}=\left(\begin{array}{cccccccc}
&& & &\lambda_1a&\lambda_1b&\lambda_1c&\lambda_1d \\
&& & & \lambda_2(-b)&\lambda_2a&\lambda_2(-d)&\lambda_2c\\
&G_n && &\vdots&\vdots&\vdots &\vdots  \\
&& & &\lambda_{2i-1}a&\lambda_{2i-1}b&\lambda_{2i-1}c&\lambda_{2i-1}d\\
&& & &\lambda_{2i}(-b)&\lambda_{2i}a&\lambda_{2i}(-d)&\lambda_{2i}c\\
& & & &\vdots&\vdots&\vdots &\vdots\\
\hline
&&&&&{\bf x}  & &\\
\end{array}\right).
\end{equation}
\end{prop}
{\pf With the above assumptions, each extended row of $G_n$, is orthogonal to itself, to $\bf x$ and to the other rows and thus the result follows. \qed}

It is clear that all rows of an $n\times n$ orthogonal matrix over ${\mathbb F}_q$ generate the ambient space ${\mathbb F}_q^n$. We can now extend to $n+2$ the dimension of the vector space generated by the rows of such an orthogonal matrix. And in this case for $q\equiv 1 \pmod 4$, we can have another construction of a self-dual $[2n+4,n+2]$ code from a self-dual $[2n,n]$ code.

\begin{prop}  Let $q\equiv 1 \pmod 4$.  Let $C_n$ be a self-dual code $[2n,n,d]$ over ${\mathbb F}_q$ with its generator matrix $(I_n|A)$. Fix $a,c\in {\mathbb F}_q$ such that $a^2\equiv c^2\equiv -1 \pmod q$.  Let ${\bf x}$ be a vector of length $n+2$ orthogonal to all extended rows of $A$ such that ${\bf x.x }\equiv 0 \pmod q.$ Then for any $\lambda_1,\hdots, \lambda_{n+1} \in {\mathbb F}_q$,  a code ${{C}'_n}$ with the following generator matrix $G_{{C}'_n}$ is a self-orthogonal $[2n+4,n+1]$ code:
\begin{equation} \label{eq:recursive3}
G_{{C}'_{n}}=\left(\begin{array}{ccc|ccccc|cc}
&&&&&&\lambda_1a&\lambda_1&\lambda_1c&\lambda_1\\
&&&&&&\lambda_2(-1)&\lambda_2a&\lambda_2(-1)&\lambda_2c\\
&I_n&&&A&&\vdots&\vdots&\vdots&\vdots \\
&&&&&&\lambda_{2i-1}a&\lambda_{2i-1}b&\lambda_{2i-1}c&\lambda_{2i-1}\\
&&&&&&\lambda_{2i}(-1)&\lambda_{2i}a&\lambda_{2i}(-1)&\lambda_{2i}c \\
&&&&&&\vdots&\vdots&\vdots &\vdots\\
\hline
&{\bf 0}&&&&&{\bf x} &&\lambda_{n+1}(-1)&\lambda_{n+1}c\\
\end{array}\right).
\end{equation}
\end{prop}
{\pf With the above assumptions, each row of $G_{{C}'_{n}}$ is orthogonal to itself and to other rows and thus the result follows. \qed}
\section{Numerical results}
In this section, we present some numerical results of good self-dual codes.
The self-dual codes are obtained either from the orthogonal matrices or from the recursive construction, we summarize some of them in Table 2 and they are available online at {\tt http://math.ahu.edu.cn/web/user.asp?id=25}\\
 The optimal, almost MDS and MDS self-dual codes together with their generator matrices and weight enumerators are arranged from smaller to larger fields. All the computations are done in Magma \cite{mag}.
\begin{table}\caption{Optimal and Best known self-dual codes}
$$
\begin{array}{|c|c|c|c|c|c|c|c|c|c|c|c|c|c|c|}
\hline
\textbf {2n/q}&\textbf {3}&\textbf {5}&\textbf {7}&\textbf {11}&\textbf {13}&\textbf {17}&\textbf {19}&\textbf {23}&\textbf {29}&\textbf {31}&\textbf {37}&\textbf {41}&\textbf {43}&\textbf {47}\\
\hline
{\bf 4}&M&A&M&M&M&M&M&M&M&M&M&M&M&M\\
\hline
{\bf 6}&&M&&&M&M& & &M& &M&M&&\\
\hline
{\bf 8}&&&M&M&M&M&M&M&M&M&M&M&M&M\\
\hline
{\bf 10}&&&&&M&M& & &M& &M&M&&\\
\hline
{\bf 12}&&A&A&M&A&6&M&M&M&M&M&M&M&M\\
\hline
{\bf 14}&&&&&7&7&&&&&7&&&\\
\hline
{\bf 16}&&&&&&8&8&&8&8&8&8&8&8\\
\hline
{\bf 18}&&&&&&&&&&&8&&&\\
\hline
{\bf 20}&&&&&&&&&&&9&&&\\
\hline
{\bf 22}&&&&&&&&&&&10&&&\\
\hline
{\bf 24}&&&&&&&10&&&&&&&\\
\hline
\hline
\hline
\textbf { 2n/q}&\textbf {53}&\textbf {59}&\textbf {61}&\textbf {67}&\textbf {71}&\textbf {73}&\textbf {79}&\textbf {83}&\textbf {89}&\textbf {97}&\textbf {101}&\textbf {103}&\textbf {107}&\textbf {109}\\
\hline
{\bf 4}&M^*&M^*&M^*&M^*&M^*&M^*&M&M^*&M^*&M&M^*&M^*&M^*&M^*\\
\hline
{\bf 6}&M&&M&&&M&&&M^*&M^*&M^*&&&M^*\\
\hline
{\bf 8}&M^*&M^*&M^*&M^*&M^*&M^*&M^*&M^*&M^*&M^*&M^*&M^*&M^*&M^*\\
\hline
{\bf 10}&M^*&&M^*&&&M^*&&&M^*&M^*&M^*&&&M^*\\
\hline
{\bf 12}&M^*&M^*&M^*&M^*&M^*&M^*&M^*&M^*&M^*&M^*&M^*&M^*&M^*&M^*\\
\hline
{\bf 14}&&&&&&&&&&&&&&7\\
\hline
{\bf 16}&&&&&&&&&&&&&&8\\
\hline
{\bf 18}&&&&&&&&&&&&&&9\\
\hline
{\bf 20}&&&&&&&&&&&&&&9\\
\hline
{\bf 22}&&&&&&&&&&&&&&10\\
\hline
\end{array}
$$
\caption*{M: MDS, A: almost MDS, $^*$: new parameters}
\end{table}

\section{Conclusion and open problems}
In this article we have constructed self-dual codes over finite fields by using orthogonal matrices.
The methods rely on the presentation of the Orthogonal group by generators and relations. A challenging conjecture on the generation
of a large subgroup of that group by transvections and permutations is presented. Another technique, closer in spirit to the extension methods of
\cite{AGKSS} is also developped.
The numerical results give more than fifty codes with new optimal parameters. Stronger machine implementation, like distributed computing
or a lower level language might lead to similar results in longer lengths.\\


\end{document}